\documentstyle[12pt]{article}
\begin{document}

\begin{titlepage}
\rightline{UTTG-28-92}
\begin{center}

 {\large \bf  Dilaton Quantum Cosmology in Two Dimensions}\\
 \vspace{1cm}
 {Francisco D. Mazzitelli\footnote{Address after
1 December 1992: Departamento de Fisica, Facultad de Ciencias
Exactas y Naturales, Universidad de Buenos Aires - Ciudad
Universitaria, Pab I - (1428) Buenos Aires, Argentina}}\\
\vspace {.3cm}
{International Centre for Theoretical Physics \\
34100 Trieste, Italy}\\
\vspace{1cm}
{Jorge G. Russo \footnote{Work supported
by NSF grant PHY 9009850 and R.A. Welch Foundation.} }\\
\vspace{.3cm}
Theory Group, Department of Physics,
University of Texas\\
Austin, Texas 78712\\

\baselineskip 18pt

\vspace{.2in}

November 1992\\
\vspace{1cm}
{\bf Abstract}
\end{center}
We consider a  renormalizable two-dimensional model of dilaton gravity
coupled to a set of conformal fields  as a toy
model for quantum cosmology.
We discuss the cosmological solutions of the model
and study the effect of including the
backreaction due to quantum corrections. As a result, when the matter
density is below some threshold  new singularities form in
a {\it weak} coupling region, which
suggests that they will not be removed in the full quantum theory.
We also solve the Wheeler-DeWitt equation.
Depending on the quantum state of the Universe,
the singularities may appear in a quantum region where the wave function is
not oscillatory, i.e., when there is not a well defined
notion of classical spacetime.
\\

\end{titlepage}

\centerline{\bf 1. Introduction}
\vskip 0.6cm
\setcounter{footnote}{0}
Two situations in which quantum gravity effects are expected to be
important are the late stages of black hole evaporation  and  the
early universe. The analysis of these problems in realistic models
is a very difficult task.
In particular, all the
attempts to address the questions of the final state of black holes,
or the evolution of the universe near Planck scale, have to deal with
the non-renormalizability of quantum gravity in 3+1 dimensions.

For this reason it is of interest to consider solvable toy models in which
some of the difficulties of the realistic problem are not present.
In the last months there has been considerable progress in the
understanding of Hawking radiation and black hole evaporation. The
fundamental observation, done by Callan, Giddings, Harvey and Strominger
\cite{cghs}
is that the 1+1 dimensional renormalizable
\footnote{Discussions on the renormalization of this model can be found
in \cite {rt}-\cite{gidd}.} theory of gravity coupled to a
dilaton field $\phi$ and $N$ conformal fields $f_i$
\begin{equation}
S_0={1\over 2\pi}\int d^2x\sqrt{-g}\left [e^{-2\phi}(R + 4(\nabla\phi)^2+
4\lambda^2) -{1\over 2}\sum_{i=1}^N(\nabla f_i)^2\right ]\ \ ,
\label{eq: model}
\end{equation}
contains black holes  and Hawking radiation.
 Some interesting discussions on this model
can be found in \cite{gidd}-\cite{bilal-callan} (for a review and more
references see ref. \cite{revi}).

In this paper we will focus on the cosmological problem.
A generalization of the  theory eq.~\ref{eq: model} to $D$
dimensions has been considered before by Vafa and Tseytlin \cite{vafa}. They
coupled
the theory to stringy matter and study the cosmological solutions
exploiting the  duality symmetry of the strings. The classical
solutions
were further investigated by Tseytlin \cite{tsey}. Our aim here  is
different. We will consider the   two dimensional theory
defined by  eq.~\ref{eq: model}  as a toy model
to understand the influence of quantum effects in cosmological
situations in higher dimensions.

In fact, a similar model can be obtained by restricting the
four dimensional action of general relativity to the metrics with
spherical symmetry:
\begin{equation}
ds^2=g_{ab}(x^0,x^1)dx^adx^b+ e^{-2\phi (x^0,x^1)}d\Omega_2\,\,\,\,
\,\,\,\, a,b=1,2 \ .
\label{eq:metric}
\end{equation}
The reduced Einstein-Hilbert action reads
\begin{equation}
S_{red}={1\over 2\pi}\int d^2x\sqrt ge^{-2\phi}\left(R+2(\nabla\phi )^2
-2\Lambda+2 e^{2\phi}\right ) \ \ ,
\label{eq:red}
\end{equation}
where $\Lambda$ is the
(four-dimensional)
cosmological constant. Note that the dilaton
field in the two dimensional theory is related to the radius of the
2-sphere in the four dimensional metric, and that the cosmological
constant gives an exponential contribution to the dilaton potential.
We will consider model \ref{eq: model} because the classical
and semiclassical analysis is simpler. Moreover, unlike the reduced
action \ref{eq:red}, the theory \ref{eq: model} is renormalizable.
Model \ref{eq: model} is analogous to
the reduced model \ref{eq:red} with $\Lambda\leq 0$. Indeed,
as far as the black hole physics is concerned, the semiclassical behaviour of
the CGHS model was shown to be very similar to the
semiclassical physics of 4D Schwarzchild black holes \cite{revi}.
On the other hand, the model given by
action \ref{eq: model} with the opposite sign in
the $\lambda^2$ term is analogous to model \ref{eq:red} with
$\Lambda>0$, i.e. standard 4D De Sitter gravity. Although in this
case there is no black hole formation in
the Universe of model \ref{eq: model},
the cosmology presents some interesting features which deserve some
attention.

An interesting particular class of spherically symmetric metrics
is the Kantowski-Sachs minisuperspace
\begin{equation}
ds^2=a^2(t)(-dt^2+dx^2)+b^2(t)d\Omega^2 \ \ ,
\label{eq:ks}
\end{equation}
which describes a Universe with a $S^1\times S^2$ spatial geometry.
This minisuperspace
has been investigated both at the classical and quantum level in ref.
\cite{laflamme}.
However, because of the non renormalizability of general relativity, it
is difficult to include quantum effects due to matter and
metric fluctuations. As we will see, it is possible to include
these effects in the  model eq. \ref{eq: model}, which can be considered
as a {\it toy} Kantowski-Sachs cosmological model after identifying
the dilaton with the radius of the 2-sphere.

The paper is organized as follows.
In the next section we discuss the exact
solutions to the backreaction problem.  That is, we  find
the time-dependent solutions of the equations
of motion which follow from the one-loop
effective action.  In Section 3 we  quantize
the effective theory.
We find different solutions to the Wheeler-DeWitt equation and
discuss whether or not they predict classical behaviour.
\vfill\eject
\centerline{\bf 2. The Back-Reaction Problem}
\vskip 0.6cm
In the conformal gauge $g_{++}=g_{--}=0$, $g_{+-}=-{1\over 2}e^{2\rho}$,
the classical action $S_0$ becomes
\begin{equation}
S_0={1\over\pi}\int d^2x[e^{-2\phi}(2\partial_+\partial_-\rho -
4\partial_+\phi\partial_-\phi +\lambda^2 e^{2\rho})
+{1\over 2}\sum_{i=1}^N\partial_+f_i\partial_-f_i]\ \ .
\label{eq: clascon}
\end{equation}
The classical equations of motion can be put into the form
\begin{equation}
\partial_+\partial_- f_i=0\ ,\ \ \ \partial_+\partial_- \rho=
\partial_+\partial_- \phi \ ,\ \ \ 2\partial_+\partial_- \phi
-4\partial_+\phi\partial_- \phi=\lambda^2 e^{2\rho}\ \ ,
\label{eq: claseq}
\end{equation}
\begin{equation}
0=e^{-2\phi}(4\partial_\pm\rho\partial _\pm \phi - 2
\partial^2_\pm \phi ) +{1\over 2}\sum_{i=1}^N\partial_\pm f_i\partial_\pm f_i
\label{eq: clasconstr}\ \ ,
\end{equation}
where we have included the equations of motion associated to the
vanishing metric components $g_{++}$ and  $g_{--}$.  The general
solution to the classical equations is \cite{cghs}
\begin{eqnarray}
f_i&=&f_{i+}(x^+)+f_{i-}(x^-)\nonumber\\
e^{-2\phi}&=&j-\lambda^2\int dx^+
\int dx^- e^{2h}\nonumber\\
\rho &=&\phi + h\nonumber\\
0&=&{1\over 2}\sum_{i=1}^N
f'_{i\pm}f'_{i\pm} + j''_{\pm}-2j'_{\pm}h'_{\pm}
\label {eq: clasol}
\end{eqnarray}
where $h=h_+(x^+)+h_-(x^-)$ and $j=j_+(x^+)+j_-(x_-)$ are free
fields.

A particular one-loop quantum effective action including the
conformal anomaly was introduced in ref. \cite{endpoint}:
\begin{equation}
S=S_0-{\kappa\over 8\pi}\int d^2x[R{1\over\nabla^2} R-2\phi R]\ \ ,
\label {eq:quact}
\end{equation}
where $\kappa ={N-24\over 12}$ is assumed to be positive.
The first (non local) quantum correction
in eq.~\ref {eq:quact} is the usual anomaly term. The second
term is local and covariant and makes the theory exactly solvable.
\footnote{In the conformal gauge this term renders the effective action
a conformal invariant field theory with vanishing central charge,
which is a consistency
requirement if the path integral is to be regularized by a naive,
field-independent  cutoff \cite{ddk} (discussions on this point
in the present context can be found e.g.
in refs. \cite{rt,bilal-callan,revi,midi}).}

By using a free field representation \cite{bilal-callan},
it is easy to find the general solution to the semiclassical equations
of motion.
We will follow closely ref.\ \cite{endpoint}.
Introducing new fields
\begin{eqnarray}
\Omega &=& {\kappa\over 2}\phi+e^{-2\phi}\ \ ,\nonumber\\
\chi &=&\kappa\rho -{\kappa\over 2}\phi +e^{-2\phi}\ \ ,
\label{eq: newfield}
\end{eqnarray}
the effective action becomes
\begin{equation}
S={1\over \pi} \int d^2x[
-{1\over \kappa}\partial_+\chi \partial_-\chi +
{1\over \kappa} \partial_+\Omega
\partial_-\Omega +\lambda^2 e^{{2\over \kappa }(\chi-\Omega)}
+{1\over 2}\sum_{i=1}^N\partial_+f_i\partial_-f_i]\ \ ,
\label {eq:newact}
\end{equation}
and the  semiclassical equations of motion are
\begin{equation}
\partial_+\partial_-\Omega=\partial_+\partial_-\chi=-\lambda^2
 e^{{2\over \kappa}(\chi-\Omega)}\ \ ,
\label{eq: bre}
\end{equation}
\begin{equation}
t_\pm=-{1\over \kappa}\partial_\pm \chi\partial_\pm\chi +
\partial_\pm^2\chi+{1\over \kappa}\partial_\pm \Omega\partial_\pm\Omega
+{1\over 2} \sum_{i=1}^N\partial_\pm f_i\partial_\pm f_i\ .
\label{eq:bred}
\end{equation}
The arbitrary functions $t_{\pm}(x^{\pm})$ reflect the
non locality of the effective action. They are fixed by the
boundary conditions necessary to define univocally the two
point function ${1\over\nabla^2}$ (see below).

The general solution
to the above equations can be written in terms of two free
fields $h$ and $j$,
\begin{eqnarray}
\Omega &=& j -\lambda^2 \int dx^+\int dx^- e^{{2\over \kappa}(h-j)}\ \ ,
\nonumber\\
\chi &=& h -\lambda^2\int dx^+\int dx^- e^{{2\over \kappa}(h-j)}\ \ ,
\nonumber\\
t_{\pm}&=&{1\over 2}\sum_{i=1}^Nf'_{i\pm}f'_{i\pm}+h_{\pm}''+
{1\over\kappa}(j'^2_{\pm}-h'^2_{\pm})\ .
\label{eq: brsol}
\end{eqnarray}
Note that
the terms proportional to $\lambda^2$ have cancelled out
in the constraint.

 From the general classical and semiclassical solutions (Eqs.~
\ref {eq: clasol} , \ref{eq: brsol}), it is easy to find time-dependent
cosmological solutions.
Let us consider coordinates $\sigma^{\pm}=\tau\pm\sigma$
and a two dimensional metric of the form
\begin{equation}
ds^2=-e^{2\rho (\tau)}d\sigma^+d\sigma^-\ \ .
\end{equation}
An  homogeneous matter distribution is $f_i=f_i(\tau )$. From
the equation of motion for the $f_i$ fields,
$\partial_+\partial_- f_i=0$, it follows $f_i=p_i \tau + b_i\ ,\ \ p_i,\ b_i=$
constants.
The functions $t_{\pm}$ depend on the quantum state of the matter fields.
The natural choice here is to impose
the matter energy momentum tensor to vanish in Minkowski
space  ($\rho (\tau)=0$).
This gives $t_{\pm}=0$ in $\sigma^{\pm}$ coordinates.
\footnote {This condition is not correct if the spatial coordinate
is compact, i.e., if $0<\sigma <L$. In this case, the matter
energy momentum tensor contains a vacuum polarization term
even when $\rho =0$. This term is given by $t_{\pm}={N\over 48 L^2}$
and will appear in the next section where the theory is quantized
on a cylinder}
{}From Eqs. (8) and (14) we find
\begin{eqnarray}
e^{-2\phi } &=& -
e^{2\lambda\tau} + 2m^2\lambda\tau +{M\over\lambda}\ ,\ \
\rho=\phi+\lambda\tau \ ,\ \ \
m^2\equiv {1\over 8\lambda^2} \sum_i p_i^2\ ,
\label{eq:omegacl}\\
\Omega &=& - e^{2\lambda\tau}+2\varepsilon \lambda\tau
+{M\over\lambda}\ ,\ \ \chi=\Omega+\kappa\lambda\tau  \  , \,\ \
\rho =\phi + \lambda\tau \ \ ,
\label {eq:omegat}
\end{eqnarray}
where $\varepsilon=m^2-{\kappa\over 4} \ $ and
the parameter $M$ is arbitrary. We have chosen
the coordinate $\tau$ in
such a way that  $\rho =\phi +\lambda\tau$ both in the
classical and semiclassical solutions.\footnote
{As
$\rho -\phi$ is a  free field we have $\rho - \phi = a\tau + b$.
The above choice is possible as long as $a$ is different from zero.
If $a=0$, $\Omega$ is quadratic in $\tau$. We will not consider
this particular solution in what follows.}

As can be seen from the structure of the propagators,
quantum loop corrections to the above solutions will
be suppressed by the coupling $g=e^{2\phi}$,
or by the effective
coupling $g^2_{\rm eff}={e^{2\phi}\over |1-{\kappa\over 4} e^{2\phi}| }$
if the path integral is performed by using the effective action which
includes the anomaly term (which implies a resummation of diagrams of
standard perturbation theory, see e.g. \cite{rt}).
Thus solutions \ref{eq:omegacl} and \ref{eq:omegat} can be trusted only
in the regions in which $g^2$ and $g^2_{\rm eff}$ are small.

Eq.~\ref{eq: newfield} implicitly defines $\phi $ in terms
of $\Omega (x^+,x^-)$.
There is a critical point
$\Omega_c ={\kappa\over 4}(1- \log{\kappa\over 4})$
 at which $d\Omega/d\phi=0$.
This trascendental equation \ref{eq: newfield} has
no solution for $\Omega < \Omega_c$, and two solutions
for $\Omega > \Omega_c$, the ``Liouville theory"
branch and the ``string theory" branch.
In the Liouville branch $\phi\in (\phi_c,\infty )$, and hence
the anomaly term in the effective action dominates over the classical
``string effective action" kinetic terms. Far from $\phi_c$,
$g^2_{\rm eff}\sim {4\over \kappa}$ and hence the $1/N$ expansion is
applicable. In the ``string theory" branch $\phi\in (-\infty,\phi_c )$
so that the anomaly term is always dominated by the classical kinetic terms.
Far away from $\phi_c$, $g_{\rm eff}^2\sim e^{2\phi}$, and the weak-field
expansion is applicable.

We will now analyze the physical behaviour of classical solutions
\ref{eq:omegacl} and
semiclassical solutions \ref{eq:omegat}. We will restrict ourselves to the
case $M>0$, other cases can be studied in a similar way.

\bigskip

Let us first consider the classical solutions \ref{eq:omegacl}.
We can distinguish two cases:

\smallskip

(i) $m^2=0$: The solution describes an expanding universe,
evolving from $\tau=-\infty$ to $\tau_0$. The curvature
and the coupling $g^2=e^{2\phi}$ are regular
at $t=-\infty$, but the scale $a^2=e^{2\rho}$
 is zero. At $\tau=\tau_0$, all the curvature, the scale, and the coupling
$e^{2\phi}$ diverge. The corresponding Penrose diagram is
illustrated by fig. 1a. This geometry can also
be interpreted as the interior of a black hole.

(ii) $m^2>0$: The solution begins at a finite value of
$\tau=\tau_1$, at which
there is a curvature singularity, and the scale and the coupling $e^{2\phi}$
diverge. The Universe contracts, the coupling becomes weaker, and then
reexpands. At finite time $\tau=\tau_2 $ there is another curvature
singularity, with infinite values for the coupling and the scale as well
(fig. 1b).

\bigskip
The semiclassical solutions can be separated in three different classes:

\smallskip

(i) $\varepsilon=0,\ M>\lambda\Omega_c$: Replacing $g^2$ by $g^2_{\rm eff}$,
this case is qualitatively the same as the classical case $m^2=0$,
but the scale is finite at $\tau_0=\tau_c$
(the two branches of eq.\ref{eq: newfield} behave in fact in a similar way).

(ii) $\varepsilon>0,\ M>\lambda\Omega_c
+\lambda \varepsilon(1-\log\varepsilon)$:
This case behaves similarly as the classical case
$m^2>0$, with $g_{\rm eff}$ diverging at initial and final times $\tau_{1,2}$,
except by the fact the scale takes finite values at $\tau_{1,2}$.

(iii) $-{\kappa\over 4}<\varepsilon<0$: Now the two solutions
given by the two
branches  of eq.\ref{eq: newfield} must be separated:

\noindent -``Liouville theory" branch: At $\tau=-\infty$,
$g_{\rm eff}^2=4/\kappa$,
$a^2 \sim e^{2\alpha\lambda\tau}\ ,\alpha={4m^2\over \kappa}$, and
 $R\sim {\rm const.} \ e^{2(1-2\alpha)\lambda\tau }$; so the scale goes to zero
($m^2\neq 0$) and there is a curvature singularity if $\alpha>1/2$.
 At $\tau=\tau_c$ the curvature
is infinity, the scale $a$ is finite, but the coupling $g^2_{\rm eff}$
diverges.

\noindent -``String theory" branch: At $\tau=-\infty$, $a=0$, $R=-\infty $,
$g_{\rm eff}^2=0$. At $\tau=\tau_c$ the curvature
is infinity, the scale $a$ is finite, but the coupling $g^2_{\rm eff}$
diverges.

A disappointing new is that in both  cases
there is a curvature singularity in a region of {\it weak} coupling.
Therefore in this case
it is very unlikely that this singularity will be removed
in the full quantum theory.
These kind of singularities could only be removed by an {\it ad hoc}
restriction of allowed initial conditions (see below).

\bigskip

It is interesting to repeat the analysis for the model with the
opposite sign in the $\lambda^2$ term.
These solutions are similar
to the static solutions in the CGHS model interchanging $\sigma $
and $\tau$. There are solutions where the Universe expands,
the coupling goes to zero,
and the metric approaches to Minkowski metric as $\tau \to\infty $
(see fig. 2a, b).
In some solutions $\phi $ never reaches $\phi_c$. In particular, there
is an interesting case with
$-\kappa/4<\varepsilon<0$: at $\tau=-\infty$,
$a=0$, $\phi=-\infty$ and $R=\infty$; at $\tau=\infty$, $a=1,\
R=0$ and $\phi=-\infty$ (fig. 2a).
Also here the coupling goes to zero even in the region
near the singularity, which makes very unlikely that more quantum effects
could wash this singularity away.

\bigskip

Now we can state the more important effects of including the matter loop
correction or backreaction:

I) There is a doubling in the number of solutions, due to the presence
of the ``Liouville theory" and ``string theory" branches. All the
semiclassical solutions with $\lambda^2>0$ reach the critical point
$\phi =\phi_{c}$ that separates both branches. There is a strong
coupling singularity there.
Some of the solutions with
the opposite sign of $\lambda^2$ never reach this point, i.e.,
they are always in a weak coupling region.

II) There appears a threshold in the matter density $m^2$,
$m_0^2={\kappa\over 4}$. Above the threshold
the cosmologies are essentially similar to the classical cosmologies,
and the singularities appear in the strong coupling regime.
Below the threshold, $-1/4<\varepsilon<0$,  singularities also occur
in a weak coupling region. This suggests that they will still be present
in the full quantum theory.

\bigskip
It is important to see whether these singularities can be removed
by proper boundary conditions.
In ref. \cite{cosmic} it was shown that
time-like, naked singularities at $\phi=\phi_c$ can be removed
by reflecting-type boundary conditions on the matter energy-momentum
tensor. The analysis is simpler in ``Kruskal'' coordinates $x^{\pm}$
where $\phi =\rho$ and $t_{\pm}={\kappa\over 4x^{\pm 2}}$ (see
e.g. \cite{endpoint}). Since
$$R={4\over 1-{k\over 4}e^{2\phi}}[\lambda^2-(\nabla\phi)^2]\ , $$
it can be finite at $\phi=\phi_c$ only if
$\partial_+\Omega \big| _{\phi=\phi_c}=\partial_-\Omega \big|_ {\phi=\phi_c}
=0$. To see under which circumstances these boundary conditions
can be applied it is interesting to
consider solutions with general matter distributions $f_i$.
They are given by eq. \ref{eq: brsol}. Note that $h_{\pm}=j_{\pm}$ in
Kruskal coordinates.
Regularity of the curvature
at $\phi=\phi_c$ requires
\begin{equation}
j'_+(x^+)=-\lambda^2x^-\ ,\ \ j'_-(x^-)=-\lambda^2 x^- \ ,\ \ \
\forall x^\pm : \phi_(x^+,x^-)=\phi_c \ .
\label{eq: bc}
\end{equation}
Eq. \ref{eq: bc} defines the shape of the boundary curve
$x^+=\hat x^+(x^-)$ and implies
a relation between the left and right-moving components of the
matter energy-momentum tensor. Using eq. \ref{eq: bc} we find
$$\lambda^2 {d\hat x^+\over dx^-}=j_-''(x^-)={\lambda^4\over
j_+''(\hat x^+)}\ .$$
Here the boundary is space-like, therefore it is possible to have
a regular curvature at $\phi=\phi_c $ only if $j_\pm ''< 0$, i.e.
(see eq. \ref{eq: brsol})
\begin{equation}
{\kappa\over 4x^{\pm 2}}
< {1\over 2}\sum_{i=1}^Nf'_{i\pm}f'_{i\pm}\ .
\label{eq: spb}
\end{equation}
For homogeneous solutions, this is satisfied in the case
$\varepsilon >0 $ discussed above.
But condition \ref{eq: bc} implies
$M=\lambda\Omega_c +\lambda \varepsilon(1-\log\varepsilon)$. In this
limiting value of $M$, $\tau_1$ is equal to $\tau_2$ and there is no
cosmological evolution.

This result
can be generalized to arbitrary matter contributions. The conclusion
is that these type of boundary conditions cannot be implemented
on space-like boundaries.
In contrast, in the model with the opposite sign of
the $\lambda^2$ term this type of boundary condition can in fact be
implemented on space-like curves. This permits cosmologies free
of singularities starting at $\phi=\phi_c$, in which the Universe expands
up to $a=1$, and the curvature and the coupling constant go to zero
as $\tau\to \infty $ (see fig. 2b).

More problematic, however, is the appearance of a singularity in
a weak coupling region. To eliminate this singularity, a
possibility is to postulate that only matter densities with
$\varepsilon\geq 0$ are allowed. Then no statement could be made
about the presence of singularities at this semiclassical level,
the only remaining singularities
would be in a strong interaction region where additional
quantum effects will be important.
\vskip 1cm

\centerline{\bf 3. The Wheeler-DeWitt Equation}
\vskip 0.6cm
We will now quantize the effective theory defined by eq. \ref{eq:newact}.
In doing this we will ignore boundary effects due to the fact that the
transformation \ref{eq: newfield}  is defined for $\Omega\geq \Omega_c $.
It is
presently unclear what are the correct boundary conditions to impose
(see ref. \cite{cosmic} ).

Assuming that the spatial coordinate $\sigma$ is periodic we
can expand the fields and their derivatives as
\begin{eqnarray}
\Omega (0,\sigma )&=&\Omega_+(0,\sigma )+\Omega_-(0,\sigma)\ \ ,
\nonumber\\
\Omega_{\pm}(0,\sigma)&=&{1\over 2}\Omega_0\pm {1\over 2}P_{\Omega}
\sigma -i\sum_{n\neq 0}{1\over n}\Omega_n^{\pm}e^{\pm in\sigma}\ \ ,
\nonumber\\
\partial_{\pm}\Omega(0,\sigma )&=&\sum_{n=-\infty}^{\infty}\Omega_n^{\pm}
e^{\pm in\sigma}\ \ , \label{eq:exp}
\end{eqnarray}
with similar expressions for $\chi$ and $f_i$.
The commutation relations are ($\Omega_0^{\pm}={1\over 2}P_{\Omega}$)
\begin{eqnarray}
&[\Omega_0,P_{\Omega}]=i{\kappa\over 2}\,\, ,  \,\,\,\, [\Omega_n^{\pm},
\Omega_m^{\pm}]=-{\kappa\over 2}\delta_{n,-m}\ \ ,\nonumber\\
&[\chi_0,P_{\chi}]=-i{\kappa\over 2}\,\, , \,\,\,\,
[\chi_n^{\pm},\chi_m^{\pm}]={\kappa\over 2}\delta_{n,-m}\ \ ,\nonumber\\
&[f_{i0},P_{f_i}]=i\,\, , \,\,\,\,
[f_{in}^{\pm},f_{im}^{\pm}]=-n\delta_{n,-m}\ \ ,\label{eq:comm}
\end{eqnarray}
(we are omitting a factor $\pi$ which comes from the global factor in
the action \ref{eq:newact}).

The Virasoro generators $L_n^{\pm}$ are defined as usual as
the Fourier modes
of $T_{\pm\pm}-{\kappa\over 2}$, where $T_{\pm\pm}$ are defined
as twice the right hand side of eq.\ref{eq:bred}. Using the equations
of motion to eliminate the second time derivatives we find
\begin{eqnarray}
L_n^{\pm}&=&\sum_m\left[{2\over k}(\Omega_m^{\pm}\Omega_{n-m}^{\pm}
-\chi_m^{\pm}\chi_{m-n}^{\pm})+f_{im}^{\pm}f_{i\, n-m}^{\pm}\right]
\nonumber\\
&+&2in\chi_n^{\pm}-{\kappa\over
2}\delta_{n0}-{\lambda^2\over\pi}\int_0^{2\pi}d\sigma
e^{\mp in\sigma}e^{{2\over\kappa}(\chi -\Omega)}\ \ .\label{eq:vir}
\end{eqnarray}
Assuming the ghosts to be in their ground state $|0>_{gh}$, the
physical states must satisfy
\begin{equation}
L_n^{\pm} |{\rm phys}>=\delta_{no}|{\rm phys}>\ \ , \ \ \ \ n\geq 0\ .
\end{equation}
This set of  equations is equivalent to the functional hamiltonian
and supermomentum constraints \cite{adm}.
The nonvanishing right hand side
in the physical state condition for $n=0$
(combined with the shift $-{\kappa\over 2}\delta_{n0}$ in the Virasoro
generators)
represents the vacuum polarization
of the physical degrees of freedom on the cylinder.

The zero modes $\chi_0$ and $\Omega_0$
do not decouple from the other modes in the constraints.
To find the physical states, we will adopt here
the minisuperspace approximation, and neglect the coupling
between the modes.
It should be stressed that this is an {\it improved}
minisuperspace approximation, since we are performing the
approximation into the effective action Eq. \ref{eq:newact}, not
in the classical action Eq. \ref{eq: clascon} \cite{midi}.
Assuming
the quantum matter fields $f_i$ to be in their ground states,
the physical states are given by $|{\rm phys}>=|0>\otimes |\Psi >$,
where $|0>$ is the state annihilated by  $\chi_n^{\pm}$, $\Omega_n^{\pm}$,
and $f_{in}^{\pm}$, for $n>0$, and the state $|\Psi >$ satisfies the
Wheeler-DeWitt (WdW) equation
\begin{equation}
\left [{\kappa\over 4}{\partial^2\over\partial\chi^2_0}-
{\kappa\over 4}{\partial^2\over\partial\Omega^2_0}-{1\over 2}\sum_i
{\partial^2\over\partial f_{i0}^2}
-4\lambda^2 e^{{2\over\kappa}(\chi_0-\Omega_0)}
-\kappa -2\right ]\Psi =0 \ \ .
\label{eq:wdweq}
\end{equation}
This equation can also be obtained by linearizing the tachyon
beta function, as observed in ref. \cite{cooper} in the context
of noncritical string theory. In that paper the authors point out that
nonlinear terms may play an important role in the cosmological
constant problem.

At this point one would like to have a criteria to choose a
particular solution to the WdW equation and consider this
particular state as the ``quantum state of the Universe".
Although there are several proposals to select a wave
function \cite{hh,vil,linde} and to  extract physical
predictions from it  \cite{kuchar},
none of them is completely satisfactory. Here
we will  analyze some particular solutions
which are the two-dimensional analogues of
some of the wave functions proposed in 3+1 quantum cosmology.
We will consider
the simplest interpretation of the wave function, i.e.
we will assume that it predicts classical
behaviour only in the regions where it is  oscillatory. The classical
trajectories associated to a wave function of the form
$\Psi = e^{iS}$ are obtained through the identification
\begin{equation}
{\partial S\over\partial\chi_0}=-{2\over\kappa}\dot\chi_0\,\, , \,\,\,
{\partial S\over\partial\Omega_0}={2\over\kappa}\dot\Omega_0
\,\, , \,\,\, {\partial S\over\partial f_{i0}}=\dot f_{i0}\ \ ,
\label{eq:limclas}
\end{equation}
where the dot denotes derivative with respect to $\tau$. It is
important to notice that, in this context, this is the {\it definition}
of classical time $\tau$ in terms of the quantum degrees of freedom \cite
{banks,hartle}.

A basis for the solutions to eq. \ref{eq:wdweq} is given by
$\Psi_p =e^{i\sum p_{i}f_{i0}}\Psi_\alpha $, where
($\chi_+=\chi_0+\Omega_0\ ,\  \chi_-=(\chi_0-\Omega_0)/\kappa $ )
\begin{equation}
(\partial_+\partial_- -\alpha-4\lambda^2 e^{2\chi_-})\Psi_\alpha =0\ \ ,
\ \ \ \ \alpha\equiv{N\over 12}-{1\over 2}\sum p_{i}^2\ ,
\end{equation}
i.e.
\begin{equation}
\Psi_\alpha = e^{i\omega} \ ,\ \ \
\omega\equiv p_-\chi_+ +p_+\chi_- -{2\lambda^2\over p_-}e^{2\chi_-}\ ,
\end{equation}
with $p_+p_-=-\alpha $.
When the numbers $p_{i},p_{\pm}$ are
large and real, this wave function is rapidly oscillating and imply
classical behaviour. The classical trajectories can be found
from eq. \ref{eq:limclas}. In terms of $\chi_\pm $ this reads
$\partial_\pm S=-\dot\chi_\mp \ $. From $S=p_if_i+\omega $
we find
\begin{equation}
\chi_- =-p_-\tau \ \ ,\ \ \ f_i=p_i\tau+b_i \ \ ,\ \ \
\chi_+=-p_+\tau -{2\lambda^2\over p_-^2}e^{-2p_-\tau }+{\rm const.}
\label{eq:traj}
\end{equation}
Eq. \ref{eq:traj} reproduces the semiclassical solutions
\ref{eq:omegat} studied in the previous section. The only
difference is the vacuum polarization contribution
${N\over 12}$ contained in $\alpha$.
Fig. 3 shows the classical trajectories
in the plane $(\chi_+,\chi_-)$.

Next we consider a popular choice: the Hartle-Hawking
state $|{\rm HH}>$, defined by the euclidean path integral
with only one boundary \cite{hh}. As noted
in refs.\cite{polch,torre}, the two dimensional analogue of
this state is the $|sl(2,C)>$ state, defined by
\begin{equation}
L_n^{\pm}|sl(2,C)>=0\,\, , \ \ \ n\geq -1 \ \ ,
\label{eq:sl2c}
\end{equation}
Let us consider the case $\lambda =  0$. Taking into account that
\begin{equation}
iL_{-1}^{\pm}=\chi_{-1}^{\pm}(2+{\partial\over\chi_0})+\Omega_{-1}^{\pm}
{\partial\over\Omega_0}+
\sum_if_{i\, -1}^{\pm}{\partial\over\partial f_{i0}}+...
\end{equation}
it is easy to show that \cite {polch}
\begin{equation}
|sl(2,C)>=|0>e^{-2\chi_0}\ \ .
\end{equation}
This state is not  physical, because it satisfies a different
$n=0$ condition. However, it becomes a physical state in the limit
$N\rightarrow\infty$. Indeed, in this limit, it coincides with
\begin{equation}
|{\rm HH}>=|0>e^{-{2\over\sqrt{1-{24\over N}}}\chi_0}\label{eq:HH}
\end{equation}
which is a particular exponential solution to the WdW equation.
In fact, this corresponds to the solution $\Psi_{p_{\rm HH}}(\lambda=0)$
with  $p_{\rm HH}=\{p_i=0 ,  p_-=i\sqrt{\alpha /\kappa} ,
p_+=i\sqrt{\alpha\kappa}\} $.
This is our analogue of the Hartle-Hawking state, which is naturally
extended to the $\lambda\neq 0 $ case by the solution
$\Psi_{p_{\rm HH}} $.
Note that this wave
function is a {\it real} exponential. As a consequence it does not
predict classical behaviour. If this were the physical state
our toy universe would never exit the quantum era!

Let us now consider other class of physical states, which are
similar to the
tunneling wave functions defined in terms of the lorentzian
path integral
in Kantowski-Sachs cosmology \cite{loukhal}.
They are solutions of the form
\begin{equation}
\Psi_\alpha =J(z)\ \ ,\ \ \
z^2=-4(\chi_+ -a)(\alpha \chi_- + 2\lambda^2 e^{2\chi_-} -b)\ \ .
\label{eq:tun}
\end{equation}
Inserting the ansatz \ref{eq:tun} into the WdW equation it is
easy to show that $J(z)$ must be a linear combination of
the Bessel functions
$J_0(z)$ and $Y_0(z)$. This class of
wave functions is oscillating when $z$
is large and real. Indeed, using the asymptotic form
of the Bessel functions we find
\begin{equation}
\sqrt z J(z)\cong \alpha e^{iz}+\beta e^{-iz}\ \ , \ \ \ \ |z|>>1\ \ .
\end{equation}
i.e., in this region of the plane $(\chi_+,\chi_-)$ the wave
function is a linear combination of WKB solutions with a Hamilton-Jacobi
function $S=\pm z$.
{}From eq. \ref{eq:limclas} we see that
both WKB components of
the wave function \ref{eq:tun} are
associated with the family of classical trajectories
\begin{equation}
{\partial _-S\over\partial_+ S}=
{d\chi_+\over d\chi_-}=
{(\chi_+-a)(\alpha +4\lambda^2 e^{2\chi_-})\over
(\alpha\chi_-+2\lambda ^2 e^{2\chi_-} -b)}\ \ ,
\end{equation}
i.e.
\begin{equation}
(\chi_+-a) =C(\alpha\chi_- +2\lambda^2 e^{2\chi_-}- b)
\end{equation}
where $C$ is an arbitrary {\it negative} constant
(it must be negative because in the classical region $z^2>0$).
Fig. 4 shows these trajectories in the plane $(\chi_+,\chi_-)$.
The plane
is  divided into classical ($z^2\gg 1$) and forbidden ($z^2\leq 1$)
regions. \footnote {A more precise definition of the forbidden
region is $z^2\leq j_{01}^2$ or $y_{01}^2$, where $j_{01}^2 (y_{01}^2)$
is the first zero of the Bessel function $J_0\, (Y_0)$. For
a similar discussion see ref.\cite{page}} We see that in general
the classical trajectories
leave, at some time, the classical region.
We should also stress that, due to the relation between
$\chi_{\pm}$ and
the original variables $\phi$ and $\rho$, the above trajectories
are meaningful only in the region $\Omega >\Omega_{c}$, i.e., when
$\chi_+ >\kappa \chi_- + 2\Omega_{c}$. This is also ilustrated in
Figs. 4.

The semiclassical solutions of the previous section contain two
different classes of singularities: weak coupling singularities
at $\tau =-\infty$  ($\chi_{\pm}=-\infty$) and strong coupling
singularities at the critical point $\Omega =\Omega_c$ ($\chi_+=\kappa\chi_-
+2\Omega_c$). From Figs. 4a and 4b, and comparing with the results of
section 2, we see that
the singularities that occur at weak coupling  (see (iii))
are present in
the classical region where the wave function is oscillatory.
However, the strong coupling
singularity may be avoided in the case $\alpha >0$. Indeed,
as shown in Fig. 4c, for
particular values of the  constants $a$ and $b$ in Eq. 33,
the curve $\Omega =\Omega_{c}$ is contained completely in the
classically forbidden region.

To summarize,
we have shown extreme examples where  the whole plane is classically
forbidden (Hartle-Hawking state), or allowed (exponential solutions
in Eq. 27). We also found more interesting solutions
where there are both classically allowed and forbidden regions
(tunneling solutions).
For some wave functions, the semiclassical singularities
are present only in the forbidden
regions. It is in this way that the choice of the quantum state
may solve the problem of the singularities, they may take place
where there is not a well defined notion of space-time.

\vspace{1cm}

{\bf Acknowledgements:}

F.D.M. would like to thank Prof. Abdus Salam, IAEA and UNESCO for
financial support. J.R. wishes to thank to L. Susskind and L. Thorlacius
for valuable discussions on related matters, and  SISSA  for hospitality.

\vspace{1cm}
{\bf Note added:}
We have received a preprint in which the
Wheeler-De Witt equation is studied in the same model
in the context of black holes
\cite{danielsson} .

\vfill\eject

\centerline{\bf Figure Captions:}

\vspace{1cm}
\underline{Figure 1(a)}: Penrose diagram for the classical solution (i).
Solid and dashed lines indicate curves of constant $\tau $ and $\sigma $
respectively.

\underline{Figure 1(b)}: Penrose diagram for the classical solution (ii).

\underline{Figure 2}:  Penrose diagrams for semiclassical solutions of
the model with $\lambda^2<0$. (a) The solution starts from $\phi_c$
where, generically, there is a curvature singularity.
Natural boundary conditions
select the only solution with regular curvature at $\phi=\phi_c $.
(b) A typical solution in which $\phi $ never
reaches the critical value.

\underline{Figure 3}: Classical trajectories defined by eq. 28 (solid lines).
The dashed-dotted line corresponds to $\Omega=\Omega_c$. The
trajectories are physical above this line. (a) Trajectories for
$\alpha <0$. (b) For
$\alpha >0$

\underline{Figure 4}: The classical trajectories defined by eq. 36.
The
solid lines are the part of
the  classical trajectories contained in the classical region. The dashed
lines are the part contained in the forbidden region. Same conventions for
the line $\Omega =\Omega_c$. (a)
The case $\alpha <0,\,\, b<{\alpha\over 2} [ln(-{\alpha\over 4\lambda^2})
-1]$. (b) The case $\alpha <0$ and
$b>{\alpha\over 2}[ln(-{\alpha\over 4\lambda^2})-1]$. (c)
The case $\alpha >0$.
$\chi_-^{(0)}$ is the zero of the function $\alpha\chi_-+2\lambda^2
e^{2\chi_-}-b$. A typical case where the line
$\Omega =\Omega_c$ is completely contained in the forbidden region.


\newpage

\begin{thebibliography}{99}



\bibitem {cghs} C.G. Callan, S.B. Giddings, J.A. Harvey and
A. Strominger, Phys. Rev. D45 (1992) R1005.

\bibitem {rt} J.G. Russo and A.A. Tseytlin, Nucl. Phys. B382 (1992) 259.

\bibitem {mm} F.D. Mazzitelli and N. Mohammeddi, {\it Classical Gravity
coupled to Liouville Theory}, Nucl. Phys. B, to appear

\bibitem {otros} S.D. Odintsov and I.L Shapiro, {\it Perturbative
analysis in two dimensional quantum gravity},
Int. J. Mod. Phys. A, to appear.

\bibitem {gidd} S.B. Giddings and A. Strominger, UC Santa Barbara
preprint, UCSB-TH-92-28 (1992).

\bibitem {cham} T. Burwick, A. Chamseddine, Nucl. Phys. B, to appear.

\bibitem  {endpoint} J.G. Russo, L. Susskind and L. Thorlacius,
{\it The Endpoint of Hawking Radiation}, Phys.Rev.D , to appear.

\bibitem  {todos} T. Banks, A. Dabholkar, M.R. Douglas and
M. O' Loughlin, Phys. Rev. D45 (1992) 3607;

S.W. Hawking, Caltech preprint, CALT-68-1 (1992);

L. Susskind and L. Thorlacius, Nucl. Phys. B, to appear;

B. Birnir, S.B. Giddings, J. A. Harvey and A. Strominger,
Phys. Rev. D46 (1992) 638.


\bibitem{bilal-callan} A. Bilal and C.G. Callan, {\it Liouville
Models of Black Hole Evaporation}, Princeton preprint, PUPT-1320 (1992);

S.P. De Alwis, {\it Black Hole Physics from Liouville
Theory}, COLO-HEP-284 (1992).


\bibitem{revi} J.A. Harvey and A. Strominger, {\it Quantum Aspects of
Black Holes}, EFI-92-41 (1992).

\bibitem {vafa} A.A. Tseytlin and C. Vafa, Nucl. Phys. B372 (1992) 443.

\bibitem  {tsey} A.A. Tseytlin, Cambridge preprint, DAMTP-92-15-REV (1992).

\bibitem {laflamme} R. Laflamme and E.P. Shellard, Phys. Rev. D35 (1987)
2315;
J. Louko, Phys. Rev. D35 (1987) 3760; J. Louko and
T. Vachaspati, Phys. Lett B223, 21 (1989); A. Gangui, F.D. Mazzitelli
and M. Castagnino, Phys. Rev. D43, 1853 (1991)

\bibitem{ddk} F. David, Mod. Phys. Lett. A3 (1988) 1651;
J. Distler and H. Kawai, Nucl. Phys. B321 (1989) 509.

\bibitem {midi} F.D. Mazzitelli, Phys. Rev. D46, xxxx (1992)

\bibitem{cosmic} J.G. Russo, L. Susskind and L. Thorlacius, {\it
Cosmic Censorship in Two-Dimensional Gravity}, Texas/Stanford preprint,
UTTG-19-92, SU-ITP-92-24 (1992).

\bibitem{adm}For a recent review see C.J. Isham, {\it Conceptual and
Geometrical Problems in Quantum Gravity}, Imperial/TP/90-91/14,
Lectures presented at the 1991 Schladming Winter School.

\bibitem{cooper} A. Cooper, L. Susskind and L. Thorlacius,
Nucl.Phys. B363, 132 (1991).

\bibitem{hh}J.B. Hartle and S.W. Hawking, Phys. Rev. D28 (1983) 2960.

\bibitem{vil} A. Vilenkin, Phys Rev D37, 888 (1988)

\bibitem{linde} A. Linde, Nuovo Cimento 39, 401 (1984)

\bibitem{kuchar}K. Kuchar, {\it Time and Interpretations of Quantum Gravity},
to be published in the {\it Proceedings of the 4th Canadian Conference
on General Relativity and Relativistic Astrophysics}, eds. G. Kunstatter,
D. Vincent and J. Williams (World Scientific, Singapore, 1992).

\bibitem{banks} T. Banks, Nucl. Phys. B249, 332 (1985)

\bibitem{hartle}J. B. Hartle in {\it Gravitation in Astrophysics,
(Cargese 1986)},
proceedings of the NATO Advanced Study Institute, Cargese, France, 1896,
eds.  B. Carter and J.B. Hartle (NATO ASI Series B: Physics, Vol.156)
(Plenum, New York, 1987).

\bibitem{polch} J. Polchinski, Nucl. Phys. B324 (1989) 123.

\bibitem{torre} D. Birmingham and C.G. Torre, Class. Quantum Grav. 4 (1987)
1149.

\bibitem{loukhal} J.J. Halliwell and J. Louko, Phys. Rev. D42 (1990) 3997.

\bibitem{page}D.N. Page, Class. Quantum Grav. 7, 1841 (1990).

\bibitem{danielsson} U. Danielsson, {\it Black Hole Uncertainties},
CERN preprint, CERN-TH. 6711/92.

\end {thebibliography}
\end{document}